# Hopping models for ion conduction in noncrystals

Jeppe C. Dyre and Thomas B. Schrøder

Department of Mathematics and Physics (IMFUFA), Roskilde University, Postbox 260, DK-4000 Roskilde, Denmark.
Fax number: (+45) 46 74 30 20; E-mail address: dyre@ruc.dk

Abstract

Ion conduction in noncrystals (glasses, polymers, etc) has a number of properties in common. In fact, from a purely phenomenological point of view, these properties are even more widely observed: ion conduction behaves much like electronic conduction in disordered materials (e.g., amorphous semiconductors). These universalities are subject of much current interest, for instance interpreted in the context of simple hopping models. In the present paper we first discuss the temperature dependence of the dc conductivity in hopping models and the importance of the percolation phenomenon. Next, the experimental (quasi)universality of the ac conductivity is discussed. It is shown that hopping models are able to reproduce the experimental finding that the response obeys time-temperature superposition, while at the same time a broad range of activation energies is involved in the conduction process. Again, percolation is the key to understanding what is going on. Finally, some open problems in the field are listed.

## 1. Introduction

Ion conduction in crystals like NaCl is relatively well understood [1]. It proceeds via well-defined defects (vacancies or interstitials in pure crystals, or impurities). Ion conduction in noncrystals poses a much greater challenge to theorists [2-4]. In this paper we briefly review the random barrier model for ion conduction in disordered solids and discuss how it compares to experiment.

## 2. The random barrier model (RBM)

It should be emphasized from the outset that this model is highly simplified. The purpose of the model is not to account for differing details of ion conduction in differing solids, but the opposite: To give the simplest possible realistic model covering the overall features of experimental dc and ac ion conduction in disordered solids. The objective is to arrive at the analogue of the *ideal gas model* for ion conduction in noncrystals. How do we know that such a model exists? We don't know for certain *a priori*, but the surprisingly universal features of ion conduction in quite different disordered solids, and the fact that these features extend to electronic and polaronic conduction in an even larger class of solids, makes one optimistic in the search for an ideal gas model of ion conduction.

Let us briefly summarize the universal features one would like to understand within a simple model [5-10]:

1) The dc conductivity is Arrhenius temperature dependent.
2) The ac conductivity follows an approximate power law with an exponent smaller than 1 which, in a fixed frequency range, goes to 1 as temperature is lowered towards absolute zero.
3) The ac conductivity obeys time-temperature superposition, i.e., the same ac response is observed at different temperatures, just displaced in the log-log plot usually used. This is often referred to as ***scaling.***
4) Different solids show roughly the same ac response: (quasi)universality.

The random barrier model concerns the motion of completely non-interacting particles on a cubic lattice. Thus not only is Coulomb repulsion ignored, but so is the self-exclusion which is present in all ion conductors (i.e., the fact that there is only room for one ion in each potential energy minimum in the solid structure). This may seem completely unrealistic, but it is not difficult to arrive at the equation describing non-interacting particles by linearizing a more general master equation (see, e.g., Ref. 10 and its references).

The random barrier model (RBM) has one further simplification, namely that the potential felt by the non-interacting charge carriers has all equal minima (Fig. 1). Again this may be justified from more general principles [10] which we shall not discuss further here. Finally, the RBM assumes the ion sites are situated on a simple cubic lattice.

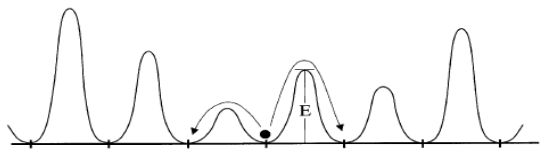

**Fig. 1:** *Typical potential for a system described by the RBM, shown here in one dimension. The barriers are assumed to vary randomly according to some probability distribution. The arrows indicate the two possible jumps for the charge carrier shown. In an external electric field the potential is tilted in one direction and a current flows.*

## 2. Three 'classical' arguments against barrier distributions

In disordered solids like glasses or polymers it seems eminently reasonable to assume that not all ion jumps involve barriers of exactly the same height. This was discussed in the literature already back in the 1950´s. However, this idea was rejected because it was thought to be inconsistent with well-established experimental facts. The following three facts were used as arguments against barrier distributions:

1) The dc conductivity is Arrhenius temperature dependent.
2) The frequency marking onset of ac conduction is Arrhenius temperature dependent with the same activation energy as the dc conductivity.
3) The ac conductivity obeys time-temperature superposition, i.e., it is possible to scale data at different temperatures to one single master curve.

The reasoning [5,11-13] based on these 3 points which apparently rules out any but an extremely narrow barrier distribution goes as follows:

1) Whenever there are barriers of differing sizes involved in the conduction process one would not expect a simple Arrhenius temperature-dependent dc conductivity. After all, which one of the many barriers involved should be chosen as the overall dc conductivity activation energy? More precisely, one should expect a non-Arrhenius behavior with smaller activation energies dominating at low temperatures where it is important to take advantage of these (perhaps relatively few), and larger activation energies gradually coming into play as temperature is increased.
2) Ac conduction must be due to ion motion over limited distances while dc conduction involves motion over extended distances (this, incidently, is absolutely correct). Consequently, one expects ac conduction to involve smaller barriers than dc conduction. And since the frequency which marks onset of ac conduction is a characteristic of ac conduction, the activation energy of this frequency must be smaller than that of the dc conductivity.
3) Any process involving a distribution of barriers must violate time-temperature superposition (unless the barrier distribution itself has a peculiar temperature dependence), because in a log-log plot the response must broaden as temperature is lowered. This is because the relevant quantity entering the transition rate is barrier divided by temperature, implying that at low temperatures more and more decades of jump frequencies are involved in the conduction process.

We now argue that these classical objections, which seem at first sight quite convincing, are in fact not valid. A more detailed discussion of this point may be found in a recent review [10].

## 3. Dc conductivity in the RBM

The first challenge to the RBM is to explain the fact that the dc conductivity is Arrhenius and to identify the activation energy. The answer is provided by percolation theory, a mathematical theory which was not invented when people in the early 1950's ruled out barrier distributions (and not known to them after 1957 when the percolation phenomenon was first discussed in the scientific literature [14]).

Suppose the barrier heights denoted by E vary randomly and uncorrelated from lattice link to lattice link of the RBM according to a probability distribution p(E). Consider the situation at low temperatures, i.e., where the barrier distribution is much broader than the thermal energy $k_B T$. We shall discuss ion motion in zero external field only, because according to the fluctuation-dissipation theorem the dc conductivity is proportional to the mean-square displacement per unit time in zero external field. Now, small barriers give rise to large jump rates, and the ions definitely prefer these. So most ion jumps proceed across small barriers. To extend the motion to infinity, however, some larger barriers obviously have to be overcome. The largest barrier which **must** be overcome to move to infinity becomes the dc conductivity activation energy. How to identify this? It is to answer this question [15,16] that percolation theory is necessary:

Consider the case of a two-dimensional simple square lattice. Suppose the links linking neighboring lattice sites are marked according to increasing barrier height. At some point in this process, referred to as the percolation threshold, an infinite cluster of marked links appear. This situation is illustrated in Fig. 2.

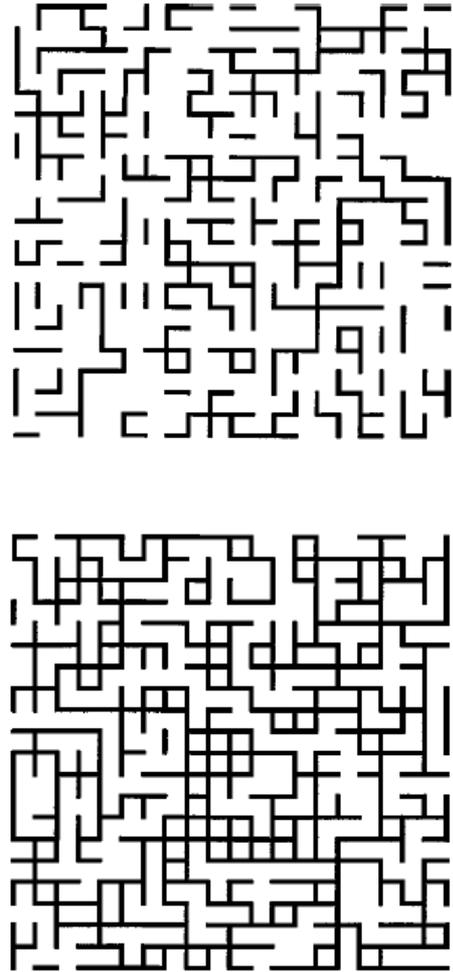

**Fig. 2:** *Percolation on a two-dimensional square lattice. The upper picture shows the situation below the percolation threshold where there is no infinite cluster of marked links, the lower picture shows the situation above the threshold. The largest barrier on the infinite cluster is the dc conductivity activation energy.*

When an infinite cluster of marked links appears (termed the percolation cluster) no larger barriers need to be overcome in order for the ion to move to infinity, corresponding to carrying a dc current. Thus, all barriers on the percolation cluster are in principle relevant for determining the dc conductivity activation energy, but no larger barriers. However, because temperature is by assumption low, the jump rates on the percolation cluster cover several decades. Consequently, the largest barrier on this cluster presents a bottleneck to the ion motion, and this largest barrier completely dominates the overall rate of motion. It is this largest barrier which becomes the dc conductivity activation energy. And it is because of the percolation phenomenon – in particular the identification of a definite bottleneck barrier – that the dc conductivity is Arrhenius temperature dependent despite the fact that a range of barriers are involved. Curiously enough, percolation theory is only relevant when a broad range of barriers is involved. So the classical argument that if there is any barrier distribution it must be quite narrow is as wrong as it could be: Only when the distribution is wide does one get an Arrhenius dc conductivity.

## 4. Ac conductivity in the RBM

What are the RBM-predictions for the ac conductivity? Figure 3 shows results of extensive computer simulations of the model. Clearly, a master curve is arrived at as the data of Fig. 3a are scaled by suitable displacements in the log-log plot (Fig. 3b), a master curve which as temperature is lowered applies in a wider and wider frequency range around the ac onset frequency. We conclude that the RBM obeys time-temperature superposition. This disproves the above classical argument 3) against barrier distributions. Our simulations [10] also show that the onset frequency has the same temperature dependence as the dc conductivity, a point we shall not dwell on here although it is an important manifestation of the celebrated Barton-Nakajima-Namikawa (BNN) relation [17-19].

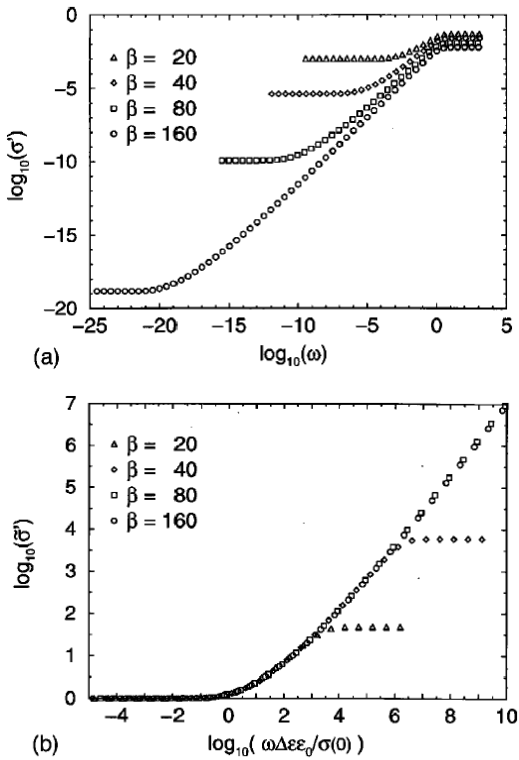

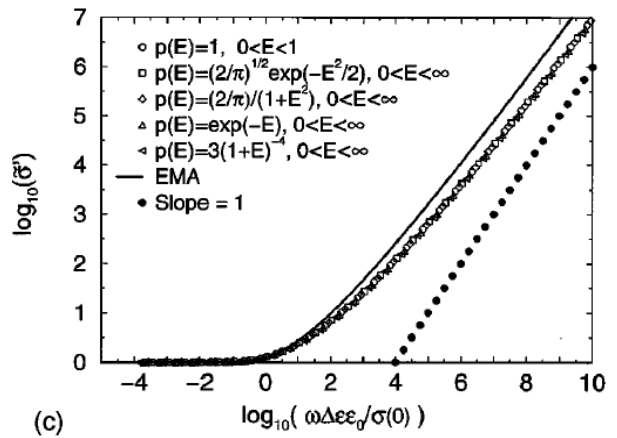

**Fig. 3:** *Computer simulations of the random barrier model in three dimensions with the uniform barrier distributions [10]. (a) shows the real part of the conductivity as function of angular frequency (in suitable units [10]) at different inverse temperatures β, (b) shows how these data may be scaled to one single master curve.*

**Fig. 4:** *Computer simulations of the random barrier model for several barrier distributions [10]. The figure shows the ac conductivity relative to the dc conductivity as function of scaled frequency at low temperatures. The dots are shown as a guide to the eye giving a line of slope one, while the full line is the classical effective medium approximation (EMA). Today more advanced analytical approximations are available [10].*

There is one further important point where the RBM reproduces experiment, namely ac universality. Figure 4 shows results of computer simulations of the RBM for several different probability distributions. This figure shows that there is universality of the ac response in the RBM. We shall not discuss the correct recipe for scaling the frequency (which was recently a subject of some debate in the scientific literature [20-23]), but just note that the onset frequency is roughly proportional to the dc conductivity in the simulations. As already mentioned, this is also seen in experiment – apparently without exception.

Close inspection of Figs. 3 or 4 reveals that the universal ac conductivity is not an exact power law as function of frequency. Rather, the ac conductivity is an approximate power law with an exponent below one, which goes slowly to one as (scaled) frequency goes to infinity. This is in agreement with experiment. In particular, we note that as temperature is lowered, by measuring in a fixed frequency range, one effectively measures further and further out on the master curve and thus finds an approximate frequency power law exponent which goes to one. This is also observed.

The frequency exponent close to one referred to above found at low temperatures or very high frequencies is usually referred to as the "nearly constant loss" (NCL). This is another universal feature which the model reproduces, but as shown in the recent literature (see, e.g., [24,25] and their references) there are other possible explanations. Nevertheless it is encouraging, we feel, that the simple RBM is able to reproduce even this feature.

To summarize the computer simulations of the RBM, the model predicts a) time-temperature superposition at low temperatures, b) universality of the ac response, and finally c) that the frequency marking onset of ac conduction has the same activation energy as the dc conductivity. How can this be understood physically? The answer is, it turns out is again, that the percolation phenomenon is responsible. It is not possible here to argue for this in detail (see [10]), but we can very briefly sketch the reasoning: Once one has established that percolation explains the Arrhenius temperature dependence of the dc conductivity as due to the fact that the dc current mainly runs on the percolation cluster, it is tempting to guess that even the ac current runs mainly on the percolation cluster. This is not quite correct, but the universal part of the ac current does indeed run on the percolation cluster [work to be published], and the universal behavior is dominated by barriers close to the dc conductivity activation energy. This fact explains not only why the onset frequency has this activation energy but also universality: The only relevant number is the value of p(E) at the dc conductivity activation energy, and even this number is "scaled away" when simulation data are scaled to arrive at the master curve. Another way of expressing this is to say that at low temperatures (in the so-called "extreme disorder limit") any barrier distribution is effectively flat. Note also that this explains time-temperature superposition: Lowering the temperature really does not change anything but the values of the jump rates (not their probabilities relative to one another).

## 5. The RBM versus experiment

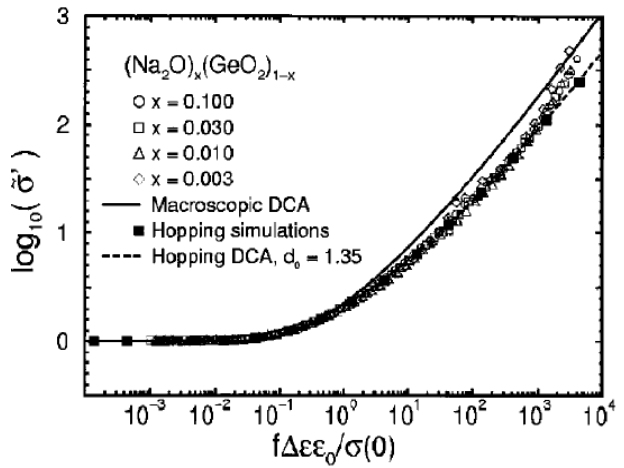

**Fig. 5:** *RBM-prediction for the ac conductivity versus data on Sodium Germanate glasses [21] with varying Sodium concentration. The open symbols are the experimental data while the full symbols give the RBM-universality prediction. "Hopping DCA" is the diffusion cluster approximation [10] (the "macroscopic DCA" is the analogue for a macroscopic model [10]).*

We have argued that the random barrier model reproduces a number of characteristic features of experimental ion conduction. But how does the model compare to experiment quantitatively? We show in Fig. 5 typical ac data compared to the universal ac conductivity of the RBM. Not all data are identical as regards the ac response, of course, so instead of referring to universality perhaps the term "quasi-universality" is more appropriate. Nevertheless, most data are reasonably well fitted by the RBM, and certainly much better than one would expect *a priori*, given the fact that the model has no fitting parameters for its prediction of the ac conductivity once model predictions are written in terms of dimensionless variables.

**6. Conclusion and open questions**

We conclude that the RBM is a simple model which captures the essential physics of dc and ac ion conduction, at least from the theoretical physicists point of view. Thus the model does indeed, in our opinion, deserve the honor of being referred to as the analogue of the ideal gas model. Nevertheless, a number of important open questions relating to the model remains to be answered:

1) How are the predictions affected when one wants to be more realistic by modifying the model to take into account Coulomb interactions and self-exclusion [3,22]?
2) How is the universal ac conductivity modified when the model is generalized to deal with sites of differing energy? Is there still ac universality?
3) How is the physical insight that the current mainly runs on the percolation cluster utilized to arrive at precise quantitative predictions of the universal ac conductivity?